\documentclass[12pt]{article}%
\usepackage{heck}
\usepackage{cite}
\usepackage{graphicx}
\usepackage{makeidx}
\usepackage{multicol}
\usepackage{geometry}
\usepackage{amsfonts}
\usepackage{mathrsfs}
\usepackage{amssymb}
\usepackage{amsmath}
\usepackage{slashed}
\providecommand{\U}[1]{\protect\rule{.1in}{.1in}}
\numberwithin{equation}{section}

\def\^{{\wedge}}
\def\*{{\star}}

\begin{document}

\date{September, 2008}

\preprint{arXiv:0809.3452}

\institution{HarvardU}{Jefferson Physical Laboratory, Harvard University, Cambridge,
MA 02138, USA}%

\title{From F-theory GUTs to the LHC}
%

\authors{Jonathan J. Heckman\footnote{e-mail: {\tt jheckman@fas.harvard.edu}}
and
Cumrun Vafa\footnote{e-mail: {\tt vafa@physics.harvard.edu}}}%

\abstract{This paper provides an overview to three recent papers on the bottom up
approach to GUTs in F-theory. We assume only a minimal familiarity
with string theory and phenomenology.  After explaining the potential
for predictive string phenomenology within this framework,
we introduce the ingredients of F-theory GUTs, and show how
these models naturally address various puzzles in four-dimensional GUT models. We next describe
how supersymmetry is broken, and show that in a broad class of models, solving the
$\mu/B\mu$ problem requires a specific scale of supersymmetry breaking consistent with a particular
deformation of the gauge mediation scenario. This rigid structure enables us to reliably extract predictions for the
sparticle spectrum of the MSSM. A brief sketch of expected LHC signals, as well as ways to falsify this class of models
is also included.}%

\maketitle

\enlargethispage{2\baselineskip}

\tableofcontents

\pagebreak

\section{Introduction\label{BOTUP}}

The landscape of semi-realistic string vacua continues to grow. This
embarrassment of riches is an obstacle in the way of extracting definite
predictions for present and future experiments. While one might have hoped that a
more complete formulation of the theory would reveal some inconsistency with
these vacua, this has proven not to be the case. Faced with this state of affairs,
extracting concrete predictions from string theory requires
further selection criteria to ascertain which of these many possibilities
could be consistent with present observations. In this regard, we believe it
is important to note that of the many known vacua of the landscape which
satisfy all global constraints imposed by gravity, not a single one
reproduces every feature of the Standard Model.

Precisely because of its success as a quantum theory of gravity, most efforts
in realizing the Standard Model within string theory have focussed on the
development of a single unified framework which couples this gauge theory to
gravity. Even so, gravity only weakly couples to the Standard Model, and can for the most part be
ignored in discussions of particle physics. At a technical level, it is also
far more cumbersome to include the effects of gravity in string based models.
For example, besides pure gravity, such models typically also contain many fields which describe
the shapes of the internal directions of a compactification. Giving all of
these moduli fields an appropriate large mass is indeed a topic of current
research, which in practice can be quite involved. In limits where gravity
decouples, many of the main issues related to moduli are also absent.

Whereas to the experimentalist, the energy scales probed by the LHC are at the
high energy frontier, to the string theorist, this is the \textit{low energy}
limit of a string compactification! Precisely because the low energy limits of
string theory are those which are most likely to confront experiment, it is
therefore natural to ignore the effects of gravity and instead focus on those
aspects of a string based construction which can replicate the correct matter
content and interactions of the Standard Model. Nevertheless, there are potentially a vast number of different
UV completions of the same IR physics. To a certain extent, this is borne
out by the existence of the string theory landscape.

The primary aim of the present paper is to argue that certain areas of the
landscape are quite constrained, and that within a bottom up approach to
string phenomenology, it is in fact possible to make definite predictions for
current and future experiments in such a framework. The essential point is
that although we may not know every detail of either the UV, or IR physics, we
may nevertheless have a crude characterization of both regimes. For example,
the IR behavior of the theory could in principle be incompatible with a given
set of UV boundary conditions. Using the renormalization group equations to
repeatedly iterate between the UV and IR, the bottom up approach can
effectively constrain the form of both sectors. Indeed, programs such as
\texttt{SOFTSUSY} \cite{SOFTSUSYAllanach} which compute the soft masses of the MSSM based
on specified UV boundary conditions adjust various inputs at the weak scale such
as $\mu$ and $B\mu$ to remain in accord with electroweak symmetry breaking.
More broadly speaking, this approach is quite commonplace in phenomenology,
but is somewhat foreign to the usual mindset adopted in string theory.

This paper provides an overview to the specific approach to string phenomenology detailed
in the three recent papers \cite{BHVI,BHVII,HVGMSB}. Combining recent insights on local model building in string
theory with elements of Grand Unified Theories (GUTs), the resulting framework turns
out to be \textit{rigid enough to extract definite predictions for phenomena seemingly
far removed from the realm of string theory}. As a general disclaimer, in
this paper we will not provide citations to related work of potential interest. We
refer the reader to the primary papers \cite{BHVI,BHVII,HVGMSB} for a more detailed list
of references and general guides to the literature.

To be sure, the aims of this approach are more modest than what was perhaps
originally envisaged in earlier bold attempts to connect string theory with
phenomenology. Indeed, by giving up on a complete description of ultraviolet
physics, it is in principle possible that important constraints could be
missed. Nevertheless, we will see that in certain cases, even in the absence
of gravity, compatibility with embedding in a string based model imposes
rather stringent conditions which are not easy to satisfy!

The rest of this paper is organized as follows. In section \ref{BASIC} we
state the basic assumptions which we shall make throughout this paper. Next,
in section \ref{sec:FINGRED} we describe the basic ingredients of F-theory
GUT\ models. In section \ref{MSSMSPEC} we show how these models realize the
exact spectrum of the MSSM. Section \ref{4dPUZZ} shows that simply achieving
the correct MSSM\ spectrum typically addresses some vexing problems of
four-dimensional GUT\ models. Supersymmetry breaking turns out to also be
quite predictive in this framework, and in section \ref{SUSYBREAK} we review
the particular deformation of gauge mediated scenarios in F-theory which solve
the $\mu/B\mu$ problem. The precise region of MSSM parameter space and a discussion of
how this class of models can be ruled out, or partially verified at the LHC is presented
in section \ref{ParamSpace}. Section \ref{CONC} contains our conclusions and
potential avenues of further investigation.

\section{Basic Assumptions\label{BASIC}}

Before proceeding to an overview of F-theory based models, we first spell out
in greater detail the main assumptions which we shall make. The first
assumption is perhaps the most crucial for bottom up string phenomenology:

\begin{itemize}
\item Low scale four-dimensional $\mathcal{N}=1$ supersymmetry is present at
energy scales which can be probed by the LHC. Moreover, this should be
interpreted as evidence for the MSSM.
\end{itemize}

Within the MSSM, the unification of the coupling constants at an energy scale
$M_{GUT}\sim3\times10^{16}$ GeV also provides circumstantial evidence for the
hypothesis of Grand Unification. Moreover, the chiral matter content of the
Standard Model organizes into three copies of the $\overline{5}\oplus10$ of
$SU(5)$, which can further unify to the $16$ of $SO(10)$ once right-handed
neutrinos are included. For these reasons, we shall also make the additional
assumption:

\begin{itemize}
\item At high energy scales, the matter content of the MSSM\ unifies into a GUT.
\end{itemize}

By this we do not necessarily mean that the resulting GUT will admit a
four-dimensional interpretation. Indeed, in all known string theory based GUT
models, the effective theory descends from a higher-dimensional gauge theory description.

Traditionally, the proximity of the GUT scale to the Planck scale $M_{pl}%
\sim10^{19}$ GeV has been interpreted as circumstantial evidence that a
further unification will occur where the gauge theory degrees of freedom unify
with gravity, perhaps through some model based on string theory. Note,
however, that there is in fact an imperfect alignment between $M_{GUT}$ and
$M_{pl}$. Indeed, $M_{GUT}/M_{pl}\sim10^{-3}$ is a small parameter, and from
the perspective of GUT\ model building, this is perhaps a quite welcome
feature! Indeed, the small value of this ratio is also consistent with
the conjecture that in any quantum theory of gravity, gravity is
the weakest force \cite{WeakGravConjecture}. At a practical level, if $M_{GUT}$ had turned out to be close to
$M_{pl}$, there would be no regime of validity for effective field theory at
the GUT scale. Moreover, in minimal incarnations of GUT models, the total
amount of matter in the theory is sufficiently small that the gauge coupling
of the GUT group is asymptotically free. In principle, then, this field theory
admits an ultraviolet completion which does not require gravity. For these
reasons, we shall also require that:

\begin{itemize}
\item There exists a limit where in principle, $M_{GUT}/M_{pl}\rightarrow0$.
\end{itemize}

We note that for realistic purposes, such a limit should not be taken, because
gravity has certainly been observed. Nevertheless, one of the perhaps
surprising outcomes of the recent work developed in \cite{BHVI,BHVII,HVGMSB} is that
simply requiring the existence of a GUT as well the existence of a decoupling
limit severely restricts the ultraviolet behavior of the theory. In fact,
\textit{pushing this framework to its logical ends will allow us to make contact with the LHC!}

\section{Elements From F-theory\label{sec:FINGRED}}

Having sketched in rough terms the bottom up approach to string phenomenology,
in this section we introduce the basic ingredients for model building in
F-theory. To this end, we now define in broad terms what is meant by
\textquotedblleft F-theory\textquotedblright, which may be viewed as a
strongly coupled formulation of type IIB string theory. In most cases, it is
assumed that the inverse coupling constant, $1/g_{string}$ assumes a constant
profile over the entire ten-dimensional spacetime. In fact, this
\textquotedblleft constant\textquotedblright, as well as its complexified
counterpart which we refer to as $\tau_{IIB}$ corresponds to the vev of a
dynamical field and as such, it can in principle have a more complicated
profile in the ten-dimensional spacetime. In F-theory, this more general case
is encoded in terms of a twelve-dimensional geometry. In addition to the four
usual large spacetime directions, this includes six internal spatial
directions as well as two additional directions which parameterize the value
of the real and imaginary parts of $\tau_{IIB}$ as they vary from point to
point in the internal directions.

The basic ingredient for model building in F-theory is a spacetime filling
seven-brane which wraps a four-dimensional internal subspace of the six
internal directions of the compactification. Such seven-branes can in general
form intersections over two-dimensional Riemann surfaces, and can also form
triple intersections at points of the internal geometry. Each lower-dimensional
subspace provides an important model building element, as in the following table:%
\begin{equation}%
\begin{tabular}
[c]{|l|l|}\hline
Dimension & Ingredient\\\hline
10d & Gravity\\\hline
8d & Gauge Theory\\\hline
6d & Matter\\\hline
4d & Yukawa Couplings\\\hline
\end{tabular}
\
\end{equation}
We now explain in greater detail the origin of the bottom three entries of this table.

\subsection{Higher-Dimensional Gauge and Matter Content}

Near a seven-brane, the profile of $\tau_{IIB}$ becomes singular. There are
only a few distinct ways that $\tau_{IIB}$ can become singular near a
seven-brane which have been found by mathematicians to be in correspondence
with the $ADE$ Lie groups $SU(N)$, $SO(2N)$, and $E_{6}$, $E_{7}$ and $E_{8}$.
The crucial point for model building is that this singularity type also
corresponds to the gauge group of the seven-brane! We note that depending on
the details of the geometry, it is also possible to engineer
non-simply-laced gauge groups such as $USp(2N)$ and $SO(2N+1)$. This flexibility in achieving a wide range of
different gauge groups in F-theory stands in contrast to the more limited
possibilities available in perturbative type IIA and IIB models where $g_{string}$ is infinitesimal. Indeed,
such models can only accommodate $SU,SO$ and $USp$ type gauge groups.

At low energies, a spacetime filling seven-brane which wraps a
four-dimensional subspace $S$ gives rise to a four-dimensional gauge
theory with coupling constant:%
\begin{equation}
\frac{1}{g_{YM}^{2}}\sim M_{\ast}^{4}\cdot Vol(S)\text{.} \label{gaugecoup}%
\end{equation}
where $M_{\ast}$ is a characteristic mass scale which relates the volume of
the six-dimensional internal space $B$ to the Planck scale via the relation:%
\begin{equation}
M_{pl}^{2}\sim M_{\ast}^{8}\cdot Vol(B)\text{.} \label{PLANCK}%
\end{equation}
The precise form of the effective action for a seven-brane with gauge group of
$ADE$ type has been determined in detail in section 3 and appendix C of
\cite{BHVI}. One of the main results of \cite{BHVI} is that detailed
properties of the geometry such as the ways in which the internal directions
can be deformed correspond to gauge theory quantities in the seven-brane
theory. Indeed, this tight correspondence provides strong evidence that the
resulting effective action accurately captures the local behavior of the seven-brane.

Proceeding down in dimension, the chiral matter of the MSSM\ originates from
configurations where two seven-branes with gauge groups $G$ and $G^{\prime}$
intersect. In such configurations, additional light degrees of freedom
described by six-dimensional fields localize along these intersections. In
terms of four-dimensional $\mathcal{N}=1$ superspace, these six-dimensional
fields can be described as a collection of vector-like pairs of fields
$\mathbb{X}$ and $\mathbb{X}^{c}$ labeled by points on the corresponding
Riemann surface. It can be shown on general grounds that the field
$\mathbb{X}$ transforms in a representation $R$ of the gauge group $G$ and
$R^{\prime}$ of the gauge group $G^{\prime}$. Similarly, $\mathbb{X}^{c}$
transforms in the complex conjugate\footnote{More precisely, the fields in
$\mathbb{X}^{c}$ transform in the dual representation to $\mathbb{X}$.}
representation $(\overline{R},\overline{R}^{\prime})$. The representations $R$
and $R^{\prime}$ are completely determined by the profile of $\tau_{IIB}$ near
the intersection locus of the two seven-branes. Starting from a gauge theory
with gauge group $G_{\Sigma}$ along the six-dimensional Riemann surface, the
basic point is that off of this curve, the original theory is Higgsed down to
either the gauge group $G$, or $G^{\prime}$, depending on which seven-brane
occupies the same location. The adjoint representation of $G_{\Sigma}$ decomposes into irreducible
representations of the subgroup $G\times G^{\prime}$ as:%
\begin{align}
G_{\Sigma}  &  \supset G\times G^{\prime}\\
ad(G_{\Sigma})  &  \rightarrow\left(  ad(G),1\right)  +\left(  1,ad(G^{\prime
})\right)  +(R,R^{\prime})+(\overline{R},\overline{R}^{\prime})+...
\end{align}
This analysis of local Higgsing in the higher-dimensional theory shows that we
should expect light degrees of freedom in the representation $(R,R^{\prime})$
localized on the Riemann surface.

In perturbative type IIB string theory, the limitations on the types of gauge groups also applies to the available types
of matter content, which must descend from the adjoint representation of
either a $SU$, $SO$ or $USp$ type gauge group. This has the important
consequence that \textit{the resulting representations always have two tensor indices.}
In F-theory, the adjoint representations of the $E$-type groups provide a small but
important set of additional possibilities. As a simple example, note that
the adjoint representation of $E_{6}$ decomposes into irreducible
representations of the subgroup $SO(10)\times U(1)$ as:%
\begin{align}
E_{6}  &  \supset SO(10)\times U(1)\\
78  &  \rightarrow45_{0}+1_{0}+16_{-3}+\overline{16}_{+3}%
\end{align}
where the $16$ is the spinor representation of $SO(10)$. Similar examples show
that other common building blocks of GUT models such as the $27$ of $E_{6}$ are also
available. As hinted at above, the role of an $E$-type
singularity is the major reason for these additional possibilities. Note, however, that
the available matter content is still rigidly determined as a descendant of an adjoint representation.
For example, in order for an $SO$ gauge group to admit massless matter in the spinor representation,
it must embed as a subgroup of $E_{8}$.

The detailed form of the effective action for a configuration of two
intersecting seven-branes may be found in subsection 4.2 and appendix D of
\cite{BHVI}. To test the form of this effective action, one can consider a
larger class of geometric deformations which describe compactifications of
F-theory in more elaborate geometries. As shown in \cite{BHVI}, in all
cases, the degrees of freedom in the intersecting seven-brane configuration
exactly match to geometric quantities in the F-theory compactification. The
detailed form of these consistency checks are presented in subsection 4.3 and
appendix F of \cite{BHVI}. We emphasize that in general, these deformations
are highly non-trivial. For example, parameterizing the form of one such
deformation fills an entire page in \cite{KatzMorrison}.

\subsubsection{Massless Matter Content\label{Massless}}

The massless particle content is obtained by expanding the higher-dimensional
field theory about a given background field configuration. In general, massless modes
can originate from both eight-dimensional fields associated with
the worldvolume of a seven-brane, or from six-dimensional fields localized at
the intersection of seven-branes. The wave functions in the internal
directions satisfy a wave equation which is of the schematic form:%
\begin{equation}
\left(  \partial+A\right)  \psi=0
\end{equation}
in either the two dimensions spanned by the Riemann surface, or the four
internal dimensions wrapped by a seven-brane. In the above, $A$ is the
background gauge field, and $\psi$ is the corresponding wave function.
Although in the constructions of \cite{BHVII} all of the chiral matter of the
MSSM\ localizes on Riemann surfaces, it is in principle possible that
some of this matter could originate from eight-dimensional fields in the
seven-brane. The precise zero mode content from eight-dimensional fields is
given in subsections 3.3.1 and 3.3.2, and the zero mode content of
six-dimensional fields is derived in subsection 4.4.1 of \cite{BHVI}. As an
explicit example, we can view the modes originating from $\mathbb{X}$ as
left-moving particles on the Riemann surface, and the zero modes from
$\mathbb{X}^{c}$ as right-moving particles. The net number of left-movers
minus right-movers is then given by the usual result available in many
textbooks:%
\begin{equation}
n_{L}-n_{R}=\underset{\Sigma}{\int}F_{\Sigma} \label{indexcomp}%
\end{equation}
where in the above, $F_{\Sigma}$ denotes the background gauge field strength
on the Riemann surface $\Sigma$.

\subsection{Interaction Terms\label{IntTerms}}

Gauge invariance of the higher-dimensional theory descends to the usual
condition that gauge fields interact with the chiral matter of the MSSM. On
the other hand, there are also additional interaction terms in the MSSM\ which
originate from the superpotential of the MSSM. In F-theory, fields localized
on Riemann surfaces will yield a cubic superpotential term when the
corresponding wave functions form a triple overlap in the internal directions.
The geometric condition for this to occur is that the corresponding Riemann
surfaces should all meet at a point in the geometry. Near this point, the
profile of $\tau_{IIB}$ becomes even more singular. Indeed, we can effectively
treat this more complicated system as an eight-dimensional gauge theory of
higher rank which locally Higgses to lower rank along each Riemann surface,
and even lower rank along the worldvolumes of the various seven-branes. In
this way, the precise form of the resulting interaction terms was obtained in
section 5 of \cite{BHVI}.

The flexibility in achieving various chiral matter representations also
extends to interaction terms. For example, in perturbative type IIB string theory, there
is a well-known obstruction to engineering the cubic coupling $5_{H}\times
10_{M}\times10_{M}$ in an $SU(5)$ GUT. Here, the subscripts indicate whether
the GUT field is a Higgs or chiral matter field. The reason this interaction
term is perturbatively forbidden is that the actual gauge group which is
perturbatively realized is $U(5)$ rather than $SU(5)$. As a consequence, this
interaction term violates the net $U(1) \subset U(5)$ of the gauge group. This
difficulty can again be traced to the limited ways in which the profile of
$\tau_{IIB}$ can change over points in perturbatively realized configurations.

When the value of the string coupling deviates away from the strict $g_{string} \rightarrow 0$ limit,
a far greater range of possibilities are available which again underscores the necessity of passing to F-theory where
$g_{string}$ is not required to be small. In subsection 5.3 of \cite{BHVI}, it is shown that
when the bulk gauge group $SU(5)$ enhances at points of the four-manifold
to the singularity $E_{6}$, the low energy superpotential contains the standard cubic term
$5_{H}\times10_{M}\times10_{M}$ of an $SU(5)$ GUT.\footnote{In more perturbative treatments of
GUT models, non-perturbative effects in $g_{string}$ are often invoked as a means to generate such
interaction terms. Note that this already requires $g_{string}$ to be an order one number.} Similarly, other common
interaction terms such as the $27^{3}$ in $E_{6}$ GUT models can also be
achieved from local enhancements from $E_{6}$ to $E_{8}$.

\subsection{Geometric Meaning of the Decoupling Limit}

One of the important features of the above ingredients is that whereas gravity
propagates in ten dimensions, in F-theory based models, all gauge theory ingredients
localize on subspaces of the compactification. We now use
this fact to sharpen the meaning of the \textquotedblleft decoupling
limit\textquotedblright described in broad terms in section \ref{BASIC}.
Geometrically, gravity can decouple when some of the internal directions
expand to infinite size. Equations (\ref{gaugecoup}) and (\ref{PLANCK}) imply that
the gauge dynamics of the MSSM do not decouple in a limit where $Vol(B)\rightarrow\infty$,
but $Vol(S)$ remains finite. Another, way to state this condition is that the four-dimensional subspace wrapped by
the GUT model seven-brane must admit a limit where it contracts to zero size
while $B$ remains of fixed size. It turns out that this condition is quite
stringent, and there is essentially a single type of four-dimensional subspace
available which satisfies these requirements known as the \textquotedblleft
del Pezzo eight surface\textquotedblright. Here, \textquotedblleft
surface\textquotedblright\ refers to a subspace which has two complex
dimensions. For further review on del Pezzo surfaces, we refer the reader to
appendix A of \cite{BHVI} and section 8 of \cite{BHVII}.

Requiring a geometric decoupling limit is also in accord with expectations based on moduli stabilization.
Insofar as the volume of the surface $S$ is stabilized by Planck scale physics,
it is natural to expect $M_{GUT}$ to be perhaps not too far from $M_{pl}$. However,
if $S$ had not been contractible, there would have been an obstruction to
making $M_{GUT}$ smaller than $M_{pl}$. This provides additional motivation
for assuming the existence of a decoupling limit. In actual
applications to GUT models, note that the value of the fine structure constant
$\alpha_{GUT}=g_{GUT}^{2}/4\pi\sim1/25$. As a consequence, $Vol(S)$ must also be
large, although still finite.

The relevant length scales of the compactification are crudely characterized by the radii
$R_{S} \equiv Vol(S)^{1/4}$, $R_{B} \equiv Vol(B)^{1/6}$, as well as a measure of the
distance scale normal to the seven-brane, $R_{\bot}$. As estimated in section 4 of \cite{BHVII}, the corresponding energy scales are:
\begin{eqnarray}
\frac{1}{R_{S}} &\sim &M_{GUT}\sim 3\times 10^{16}\text{ GeV} \\
\frac{1}{R_{B}} &\sim &M_{GUT}\times \varepsilon ^{1/3}\sim 10^{16}\text{ GeV} \\
\frac{1}{R_{\bot }} &\sim &M_{GUT}\times \varepsilon ^{\gamma_{\bot}}\sim 5\times
10^{15\pm 0.5}\text{ GeV}
\end{eqnarray}
where:
\begin{equation}\label{EPSDEF}
\varepsilon \equiv \frac{M_{GUT}}{\alpha _{GUT}M_{pl}}\sim 7.5\times 10^{-2}
\end{equation}
and $1/3 \leq \gamma_{\bot} \leq 1$ is a measure of the eccentricity of the normal directions to the four-manifold $S$. The hierarchy $M_{GUT}/M_{pl} \sim 10^{-3}$ will appear repeatedly in estimates on the axion decay constant, the masses of neutrinos, as well as the $\mu$ term. Finally, note that as required to achieve a decoupling limit:
\begin{equation}
R_S < R_{B} < R_{\bot} \text{.}
\end{equation}

\section{Achieving the Exact Spectrum of the MSSM\label{MSSMSPEC}}

Up to this point, we have simply given a broad set of general considerations
for local model building in F-theory GUTs. Here, we show
that the two conditions:

\begin{itemize}
\item A supersymmetric GUT\ exists;

\item There exists a limit where gravity can in principle decouple,
\end{itemize}

impose surprisingly powerful restrictions on the ultraviolet behavior of the
four-dimensional effective field theory. Indeed, as we explain in this
section, simply achieving the correct matter spectrum will automatically
address a number of puzzles present in four-dimensional GUT models. While any
one solution might be considered at best only circumstantial evidence for this
approach, we find it compelling that a single ingredient typically addresses
several issues in traditional four-dimensional GUT\ models simultaneously.

The starting point for our discussion is that the chiral matter content of the MSSM
should organize into representations of a GUT group. Achieving this requires
that the gauge degrees of freedom descend from a seven-brane which
contains the GUT\ group $SU(5)$. In these models, the chiral matter and Higgs
fields of the MSSM correspond to zero modes of six-dimensional fields which
localize on Riemann surfaces in the four-dimensional subspace wrapped by the
GUT\ model seven-brane. For example, in the explicit minimal realizations of
an $SU(5)$ GUT\ presented in section 17 of \cite{BHVII}, the $\overline{5}%
_{M}$, $\overline{5}_{H}$ and $5_{H}$ localize on various Riemann surfaces
where the singularity type enhances from $SU(5)$ to $SU(6)$. In addition, the
$10_{M}$ fields localize on Riemann surfaces where the singularity type
enhances to $SO(10)$.

In traditional four-dimensional GUT models, breaking the GUT group is
typically achieved by allowing an adjoint-valued chiral superfield to develop
a suitable vev in the $U(1)_{Y}$ hypercharge direction, breaking $SU(5)$ to
$SU(3)_{C}\times SU(2)_{L}\times U(1)_{Y}$. In appendix E of \cite{BHVI}, it
is shown that in models which admit a decoupling limit, no such chiral
superfields exist in the four-dimensional effective theory. As a consequence,
the usual four-dimensional GUT\ scenario cannot be realized! This feature is
also quite common to many other string based models which aim to realize the
MSSM. In that context, another common approach is to consider gauge group
breaking via Wilson lines. This requires that certain topological conditions
must be met in the internal directions of the compactification. As
explained in section 5 of \cite{BHVII}, models which admit a decoupling limit
do not satisfy this criterion. Thus, the two main ways that have been
attempted to even break the GUT\ group are simply unavailable in local
F-theory models.

It turns out that there is another way to break the GUT group which is
somewhat unique to F-theory. This corresponds to turning on an internal flux
on the GUT\ model seven-brane which is aligned in the $U(1)_{Y}$ hypercharge
direction. Given its simplicity, it may at first appear surprising that this
mechanism is not used more frequently in string based constructions. Indeed,
there is a well-known obstruction to utilizing this mechanism in contexts
other than F-theory because the corresponding field strengths couple to axion-like fields in the
four-dimensional effective theory. Via the St\"uckelberg mechanism, such couplings
end up generating a large mass for the
$U(1)_{Y}$ gauge boson, in sharp disagreement with observation.

However, as recently shown in \cite{BHVII,DonagiWijnholtBreak}, such problematic
couplings to axion-like fields are absent when the internal flux obeys a certain
set of topological conditions. For discussion on the precise form of this
condition, as well as some explicit examples where it can be satisfied, see
section 9 of \cite{BHVII}.

Usual expectations from string theory might suggest that the mechanism for GUT
group breaking detailed above is present in some dual formulation of the theory.
Indeed, there is a well known duality between certain compactifications of F-theory
and the heterotic string. In fact, the geometry of the compactification must be of
a rather special type in order for this duality to hold. As explained in section
9.1 of \cite{BHVII}, the mechanism detailed above turns out to be unavailable
in those models which possess a heterotic dual!

The presence of an internal hypercharge flux, or \textquotedblleft
hyperflux\textquotedblright\ has immediate repercussions throughout the rest
of the model. Just at the level of the spectrum, this background flux can
sometimes generate matter fields in exotic representations. The precise
conditions for avoiding such exotica are discussed in detail in section 10 of
\cite{BHVII}. One consequence of this work is that the higher the rank of the
GUT\ group, the more difficult it is to remove the exotics. In this way, it
was shown there that if all of the matter of the MSSM localizes on Riemann surfaces, the GUT group must correspond either to $SU(5)$ or
$SO(10)$ in eight dimensions. In the latter case, the corresponding model
descends to a flipped $SU(5)$ GUT model in four dimensions. One interesting feature
of this scenario is that typical problems with embedding four-dimensional
flipped GUT\ models in $SO(10)$ gauge groups are absent in this
higher-dimensional approach.

The background hyperflux also provides a conceptual explanation for why the
Higgs fields do not organize into full GUT multiplets, whereas the rest of the
chiral matter of the MSSM does. Recall that the chiral matter of the
MSSM\ descends from zero modes of six-dimensional fields which localize on
Riemann surfaces inside of the GUT\ model seven-brane. These six-dimensional
fields couple to $U(1)$ gauge fields associated with seven-branes which intersect the GUT
model seven-brane, as well as the activated hyperflux. The individual
components of a representation charged under the GUT\ group $SU(5)$ will
couple differently to the $U(1)_{Y}$ hyperflux. For example, the
$5$ of $SU(5)$ decomposes to the Standard Model gauge group as:%
\begin{align}
SU(5)  &  \supset SU(3)_{C}\times SU(2)_{L}\times U(1)_{Y}\\
5  &  \rightarrow(1,2)_{1/2}+(3,1)_{-1/3}\text{.}%
\end{align}
In a suitable integral normalization of charge, this implies that the total
number of left-movers minus right-movers in the $(1,2)_{1/2}$ versus
$(3,1)_{-1/3}$ is:%
\begin{align}
(1,2)_{1/2}  &  :n_{L}-n_{R}=3\underset{\Sigma}{\int}F_{U(1)_{Y}}%
+q\underset{\Sigma}{\int}F_{U(1)_{\bot}}\label{doub}\\
(3,1)_{-1/3}  &  :n_{L}-n_{R}=-2\underset{\Sigma}{\int}F_{U(1)_{Y}}%
+q\underset{\Sigma}{\int}F_{U(1)_{\bot}} \label{trip}%
\end{align}
where in the above, $U(1)_{\bot}$ refers to a $U(1)$ from a seven-brane which
intersects the GUT\ model seven-brane, and $q$ denotes the charge of the
six-dimensional field under this gauge group. As is apparent from equations
(\ref{doub}) and (\ref{trip}), when the net hyperflux is non-zero, the
resulting matter content cannot organize into full GUT\ multiplets.
Conversely, the zero mode content will always form GUT\ multiplets when the
net hyperflux vanishes. This yields the simple criterion:%
\begin{align}
\text{Higgs}  &  \text{:}\underset{\Sigma}{\int}F_{U(1)_{Y}}\neq0\\
\text{Matter}  &  \text{:}\underset{\Sigma}{\int}F_{U(1)_{Y}}=0\text{.}
\label{mattflux}%
\end{align}
Note that in order to achieve a chiral matter spectrum, the net flux from
the $U(1)_{\bot}$ factor must therefore be non-trivial. Explicit
examples which achieve the exact spectrum of the MSSM based on $SU(5)$ GUTs
are given in section 17, and models based on flipped $SU(5)$ GUTs are
presented in section 18 of \cite{BHVII}.

\section{Addressing Four-Dimensional GUT\ Puzzles\label{4dPUZZ}}

It may appear somewhat perplexing that our ultimate goal is a realization of a
GUT model, because there are also well-known difficulties with traditional
four-dimensional GUTs. In this section we show that issues such as proton decay, GUT mass relations,
and large masses for vector-like pairs, as well as more positive features
such as variants on the seesaw mechanism for neutrino masses are naturally addressed
in local F-theory GUTs. As we now explain, the essential reason for this is that
the $U(1)$ hyperflux which plays a prominent role in achieving the correct matter
spectrum also enters more indirectly into other aspects of the low energy physics.

\subsection{Proton Decay and Doublet-Triplet Splitting\label{DTS}}

One obstacle in realizing semi-realistic four-dimensional GUT\ models is the
so-called doublet-triplet splitting problem. In its mildest form, this is the
fact that as opposed to the chiral matter of the MSSM, the Higgs fields do not
fit into complete GUT\ multiplets. While any string based model which realizes
the MSSM must provide a solution to this problem, the fully story is more
complicated. This is because heavy Higgs triplets can still mediate proton
decay. Indeed, heavy triplet exchange could still generate the superpotential term:%
\begin{equation}
W\supset\frac{\eta}{M}QQQL\text{,}%
\end{equation}
where $Q$ denotes a quark doublet superfield, and $L$ denotes a lepton doublet
superfield. Even if $M$ is on the order of the Planck scale, the parameter
$\eta$ must be quite small in order to avoid rapid nucleon decay. This operator
is generated when the Higgs triplet associated with $H_{u}$ and
the one associated with $H_{d}$ obtain a large mass through the superpotential term:
\begin{equation}
W\supset MT_{u}T_{d} \label{tripmass}%
\end{equation}
in the obvious notation. In higher-dimensional theories, it is important to note that even if
the zero mode spectrum does not contain such fields, their Kaluza-Klein modes will still be present.

This potential problem is absent in local F-theory GUT\ models. It follows from equation (\ref{doub}) that a given matter curve will
typically only support one type of Higgs field. As a consequence, the Higgs up
and down fields must localize on distinct matter curves. Instead of
equation (\ref{tripmass}), the resulting mass terms are therefore of
the form:%
\begin{equation}
W\supset MT_{u}T_{d}^{\prime}+MT_{d}T_{u}^{\prime}%
\end{equation}
where the $T^{\prime}$'s are additional triplet chiral superfields corresponding
to Kaluza-Klein modes of the theory. Hence, the operator $QQQL$ is
not generated by heavy Higgs triplet exchange.\footnote{In
the four-dimensional GUT model building literature, this is known as the missing partner mechanism.} Further
discussion on this point, as well as a more general explanation based on the
presence of background $U(1)$ symmetries may be found in sections 12 and 13 of
\cite{BHVII}.\ See \cite{DonagiWijnholtBreak} for some additional recent
discussion on proton decay in local F-theory models.

\subsection{GUT Mass Relations}

In a strictly four-dimensional model, note that because the chiral matter of the
MSSM\ organizes into GUT multiplets, gauge invariance of the higher rank
gauge group requires that the cubic interaction terms of the superpotential must
assume a far more constrained form than what is necessary to realize the Yukawa couplings of the MSSM.
For example, the right-handed down type quarks and lepton doublets organize into the $\overline{5}$ of
$SU(5)$. As a consequence, we can expect the mass relation:
\begin{equation}
m_{b}=m_{\tau} \label{massrel}%
\end{equation}
to hold at the GUT scale. While this relation holds fairly well for the
heaviest third generation, it is violated for the lighter two generations.
Various elaborate mechanisms have been proposed to circumvent this problem
based on either including higher dimension operators in the superpotential, or
by including fields in larger dimension representations of the GUT group which can
induce further structure in the Yukawa couplings once they develop suitable
vevs. In F-theory based models, the second option is not even available.
Indeed, a rough classification of the possible representations available
from minimal rank enhancements can be found in appendix C of \cite{BHVII}.

A priori, the presence of a background hyperflux will distort these GUT\ mass
relations. As reviewed near equation (\ref{mattflux}), the \textit{net}
hyperflux through a Riemann surface supporting a chiral generation must vanish
in order for the number of irreducible components of a GUT multiplet to remain
equal. This, however, is a discrete quantity, and in particular is not
sensitive to detailed properties of the flux at individual points of the
Riemann surface. Indeed, it is a far more severe requirement to demand that
the flux vanish pointwise on the Riemann surface. Because different
irreducible representations of the Standard Model gauge group have
distinct $U(1)$ hypercharges, the resulting wave functions on the
Riemann surface will also be dissimilar. In particular, the overlap of these
wave functions with the Higgs field wave functions will in general be
different. For this reason, it would at first seem that there is no reason to
expect any relation of the type given by equation (\ref{massrel}).

At a qualitative level, this distortion also decreases as the mass of a field increases
because the kinetic term for a field localized on a Riemann surface $\Sigma$ is proportional
to its volume, $Vol(\Sigma)$. In a canonical normalization of all fields, the mass therefore scales as:
\begin{equation}
m\varpropto\frac{1}{\sqrt{Vol(\Sigma)}}\text{.}%
\end{equation}
As $Vol(\Sigma) \rightarrow 0$, the cost in energy to maintain a large
imbalance in flux increases. Hence, for smaller curves, the hyperflux also
diminishes in strength. In this case, the wave functions of different components
of a GUT\ multiplet will have similar profiles over the Riemann surface. The corresponding mass relation of equation
(\ref{massrel}) will therefore become qualitatively more accurate for heavier
fields, just as is observed. See section 14.3 of \cite{BHVII} for a slightly
more technical version of this same discussion.

\subsection{Singlet Wave Functions and Neutrino Masses\label{NEUTRINOS}}

One of the successes of the GUT paradigm is the seesaw mechanism. This
provides a qualitative explanation for why neutrino masses can in general be
far lighter than the weak scale. This is usually taken as evidence for $SO(10)$ GUTs,
and the fact that with right-handed neutrinos, the matter content of the MSSM unifies
into the spinor $16$ of $SO(10)$. In local F-theory GUT models, such spinors can
be accommodated even when the bulk gauge group is only $SU(5)$ when the bulk singularity
type enhances by more than one rank along a Riemann surface. Some related examples of this type are
presented in section 7 of \cite{HVGMSB}.

A variant of the seesaw mechanism is also available in $SU(5)$ GUTs even when
the singularity type enhances by a minimal amount. This is because singlet
fields will generically interact with lepton doublets in such a way that
they can be consistently identified with right-handed neutrinos. In these cases,
the Majorana mass of the heavy neutrinos is typically somewhat lighter than in a traditional four-dimensional
GUT\ model. Recall that in the MSSM, the lepton doublet and Higgs up field
form a vector-like pair. As explained in section 15.1 of \cite{BHVII},
vector-like pairs of fields $\rho$ and $\rho^{\prime}$ can interact with a
$SU(5)$ GUT group singlet $X_{\bot}$ through an interaction term of the form:
\begin{equation}
W\supset\kappa X_{\bot}\rho\rho^{\prime}\text{.}%
\end{equation}
As a singlet, the Riemann surface supporting $X_{\bot}$ does not reside inside
of the GUT model seven-brane, but rather will only intersect it at distinct
points. In particular, this implies that the behavior of the $X_{\bot}$ wave
function will behave in a qualitatively different fashion from $\rho$ and
$\rho^{\prime}$. This singlet wave function was analyzed in detail in
subsections 15.1 and 15.2 of \cite{BHVII} where it was shown that the local
positive curvature of the four-dimensional surface wrapped by the seven-brane
can either repel or attract the $X_{\bot}$ field wave function away from the
GUT\ model seven-brane. When it is attracted towards the seven-brane, $\kappa$
is typically suppressed by a small overall volume effect related to the small
hierarchy of scales $M_{GUT}/M_{pl}\sim10^{-3}$. On the other hand, when
$X_{\bot}$ is repelled away from the seven-brane, $\kappa$ will be
exponentially suppressed, providing a potential mechanism for generating
large hierarchies in energy scales. As specific
applications, it was proposed in \cite{BHVII} that this type of suppression
term could generate a very small $\mu$ term, or light Dirac neutrino masses.
Some potential drawbacks are that this type of exponential suppression
provides a mostly qualitative picture, which is in itself not very predictive.

In the more predictive case where $\kappa$ is only suppressed by a mild volume
related factor, it can be shown that the superpotential for the neutrinos is
of the rough form:%
\begin{equation}
W=\alpha_{GUT}^{3/4}\left(  H_{u}LN_{R}+M_{GUT}\varepsilon^{4\gamma_{\bot}}\cdot
N_{R}N_{R}\right)  \text{.}%
\end{equation}
where $L$ denotes the lepton doublet, $H_{u}$ the Higgs up and $N_{R}$ the
right-handed neutrino chiral superfield. Here, $\varepsilon$ is the same
volume suppression factor defined by equation (\ref{EPSDEF}), and
$1/3 \leq \gamma_{\bot} \leq 1$ is again a measure of the eccentricity
in directions normal to the GUT model seven-brane.
The resulting value for the Majorana masses of the right-handed neutrinos is
typically close to $10^{12}$\ GeV, and the light neutrinos have mass:
\begin{equation}
m_{\text{light}}\sim\alpha_{GUT}^{3/4}\varepsilon^{-4\gamma}\frac{\left\langle
H_{u}\right\rangle ^{2}}{M_{GUT}}\sim2\times10^{-1\pm1.5}\text{ eV.}%
\end{equation}
This provides a small enhancement over the usual value $\left\langle
H_{u}\right\rangle^{2}/M_{GUT}$ of the simplest GUT based seesaw mechanism.
This value is in slightly better accord with expectations based on neutrino oscillations.
Further discussion on neutrinos in local F-theory models may be found in
subsection 15.4 of \cite{BHVII}. Of course, this
analysis should only be viewed as a first step towards a more complete theory of
neutrinos.

\subsection{Towards a Theory of Flavor}

Flavor physics is an important component of any model which
aims to incorporate the Standard Model which has so far met with limited success
in string based models. The interplay between the geometry of the matter curves and
the Yukawas of the four-dimensional effective theory is described in section 14 of \cite{BHVII}.
At zeroth order, the most important requirement is that the top quark should
have much larger mass than the other quarks of the Standard Model. The geometric
conditions for semi-realistic textures are described in subsections 14.1 and 14.2 of \cite{BHVII}.

Some general speculations on realizing a hierarchical CKM matrix via a geometric
realization of the Froggatt-Nielsen mechanism
are presented in subsection 14.4 of \cite{BHVII}. Additional speculations on possible
discrete flavor symmetries originating from the symmetry groups of del Pezzo surfaces
are presented in subsection 14.5 of \cite{BHVII}.

\subsection{Addressing the Crude $\mu$ Problem\label{CRUDEMU}}

Achieving the correct matter spectrum has another important consequence in the
Higgs sector of the theory. In subsection \ref{DTS} we have emphasized the
role of the heavy Higgs triplets in doublet triplet splitting. On the other
hand, it is also important that the Higgs doublets remain light. As before,
the main point is that achieving the correct matter spectrum requires that the
Higgs up and Higgs down fields must localize on different Riemann surfaces.
This already addresses the crudest version of the $\mu$ problem which is the
puzzling fact that in the MSSM, the vector-like pair of Higgs fields could develop a large
mass through the superpotential term:%
\begin{equation}
W\supset\mu H_{u}H_{d}\text{.}%
\end{equation}
In local F-theory constructions, the Higgs fields originate at the
intersection of the GUT model seven-brane with other seven-branes. This
implies that the Higgs fields are charged under additional $U(1)$ gauge group
factors which forbid bare $\mu$ terms. When the Higgs curves do not intersect,
the $\mu$ term is absent. When these curves do meet, the Higgs fields could interact with a
GUT group singlet $X$, either through an F- or D-term. These contributions can induce
an effective $\mu$ term once $X$ develops a suitable vev. In principle, when the
resulting singlet wave function is exponentially suppressed, very small values for the
$\mu$ term are also possible. In the next section we will address a more
refined version of this same issue where the value of $\mu$ correlates with the
scale of supersymmetry breaking.

\section{Supersymmetry Breaking\label{SUSYBREAK}}

One of the main themes of this paper is that imposing only a few qualitative
assumptions on the behavior of F-theory GUT models is enough to tightly
constrain the ultraviolet behavior of the effective field theory. Even so, any
semi-realistic model which aims to incorporate the MSSM must also address the
origin of supersymmetry breaking. In most viable scenarios, there are three
sectors corresponding to the visible sector, the hidden sector where
supersymmetry is broken, and the messenger sector which communicates these
effects to the visible sector. In most models, the mediation sector proceeds
either through Planck suppressed operators, as in gravity mediation,
or through the gauge fields of the Standard Model, as in gauge
mediation, and there are many variants on these two basic possibilities. In this
section we review some of the analysis of \cite{HVGMSB} showing that crude
considerations based on correlating the weak scale with the scale of
supersymmetry breaking determine to a remarkable extent the IR behavior of this class of
models. In fact, this information is sufficiently precise that in section
\ref{ParamSpace} we will be able to determine the primary characteristics of
the sparticle spectrum for a broad class of F-theory GUTs.

\subsection{Addressing the Refined $\mu/B\mu$ Problem}

Correlating the scale of supersymmetry breaking with
the weak scale requires that the $\mu$ term should somehow be sensitive to the
effects of supersymmetry breaking. Parameterizing the effects of supersymmetry
breaking by a chiral superfield $X$ which develops a supersymmetry breaking
vev:%
\begin{equation}
\left\langle X\right\rangle =x+\theta^{2}F\text{,} \label{XVEV}%
\end{equation}
the scale of supersymmetry breaking in the hidden sector is given by $\sqrt
{F}$. In order to spontaneously break supersymmetry, $X$ should be treated as
a dynamical field rather than as a spurion field, and we shall therefore only
consider this possibility. In this section we demonstrate that there are a broad class of vacua in
F-theory where $\mu(F,\overline{F})$ is a non-constant function. One of the
main results of \cite{HVGMSB} is that the resulting value of the $\mu$ term
typically requires $F\sim10^{17}$ GeV$^{2}$ to solve the $\mu$ problem. This
turns out to imply that gravity mediation would generate a value for $\mu$
far above the weak scale. Assuming that supersymmetry breaking
communicates to the visible sector via gauge mediation, this also requires
that $x\sim10^{12}$ GeV.

These crude restrictions stem from requiring that the vev of $X$ determines
$\mu(F,\overline{F})$. As implicitly assumed throughout \cite{HVGMSB}, this is
naturally realized when the wave function for $X$ overlaps with the wave
functions for the Higgs fields. Matter fields which localize on Riemann
surfaces interact most strongly at points of maximal wave function overlap.
Geometrically, this requires that the corresponding Riemann surfaces touch at
some point. As reviewed in subsection \ref{IntTerms}, when two matter
curves intersect, there will always be a third matter curve which also touches
at this same point.\footnote{This follows by analyzing the local profile of
$\tau_{IIB}$ near such a point of intersection.} For these reasons, it is
perhaps most straightforward to assume that $X$ localizes on a Riemann surface
which forms a triple intersection with the matter curves supporting the Higgs
up and Higgs down fields.

In this case, section 4 of \cite{HVGMSB} establishes that either the following
F- or D-term will be present in the low energy theory:
\begin{equation}
\int d^{2}\theta XH_{u}H_{d}\text{ \ \ or }\int d^{4}\theta\frac{X^{\dag}%
H_{u}H_{d}}{M_{X}}%
\end{equation}
where the $U(1)$ symmetries of the background seven-branes prevent both
terms from appearing in the same action. In the above, the D-term is
generated by integrating out Kaluza-Klein modes of the $X$ field
which have mass $M_{X}\sim 10^{15.5}$ GeV.

First consider the case where the F-term is allowed. If present, this operator would
exacerbate the $\mu/B\mu$ problem for typical
values of $x$ and $F$. Moreover, $\mu$ would depend on neither $F$ nor $\overline{F}$.
On the other hand, the D-term is potentially
more promising for generating a potential correlation because it realizes a
variant of the Giudice-Masiero mechanism \cite{GiudiceMasiero}. When $X$
develops a supersymmetry breaking vev, the D-term will induce an effective $\mu$ term with:%
\begin{equation}
\mu\sim\gamma\frac{\overline{F}}{M_{X}}\text{.}%
\end{equation}
In practice, $\gamma$ is typically an order $10$ number so that for
$F\sim10^{17}$ GeV$^{2}$, $\mu$ is close to the weak scale. In gravity
mediated scenarios, $F \sim 10^{21} - 10^{22}$ GeV$^{2}$, which would generate
too large a value for $\mu$ in the present class of models.
Finally, the phenomenological requirements of gauge mediation also imply $F/x\sim10^{5}$ GeV
so that $x\sim10^{12}$ GeV. When we discuss the explicit supersymmetry breaking scenario
based on a Fayet-Polonyi model, we will show that this value for $x$ naturally emerges
from string based considerations.

It may at first seem somewhat perplexing that this natural mechanism has not
previously been more exploited in the phenomenology literature. In a generic
effective field theory, the suppression scale $M_{X}$ is naturally identified with $x$.
This would not solve the $B\mu$ problem, however, because the D-term:
\begin{equation}
\int d^{4}\theta\frac{X^{\dag}XX^{\dag}H_{u}H_{d}}{M_{X}^{3}}%
\end{equation}
generates a $B\mu$ term when $X$ develops a supersymmetry breaking vev:
\begin{equation}
B\mu\sim\frac{\overline{x}\left\vert F\right\vert ^{2}}{M_{X}^{3}}\text{.}%
\end{equation}
Hence, when $x\sim M_{X}$, we find
$B\mu\sim \left\vert F/x \right\vert^{2} \sim (10^{5}$ GeV$)^{2}$ at the messenger scale,
which is problematic. Note, however, that in
the present case, the value of $B\mu$ is far smaller at the messenger scale
because $x/M_{X}\sim10^{-3}$, solving the $B\mu$ problem. For further
discussion on this point, see section 4 of \cite{HVGMSB}. An
explicit realization of the gauge mediation scenario which solves the
$\mu/B\mu$ problem through the described variant of the Giudice-Masiero
mechanism was constructed in section 5 of \cite{HVGMSB} and is referred to as
the `diamond ring model'.

At the messenger scale, $B\mu = 0$, and all of the $A$-terms of the soft
supersymmetry breaking Lagrangian also vanish. One consequence of this is
that the argument of all of these terms at lower scales are correlated, thus
preventing any extraneous CP violation in the low energy Lagrangian. This point
was already briefly mentioned in section 16 \cite{BHVII} and was implicitly
assumed throughout \cite{HVGMSB}.

\subsection{Consequences of an Anomalous $U(1)$ Peccei-Quinn Symmetry}

\subsubsection{$E_{6}$ Embedding}

From the perspective of the effective field theory, the bare $\mu$ and $B\mu$
terms are forbidden by requiring that all fields of the MSSM have appropriate
charges under a $U(1)$ Peccei-Quinn symmetry. In local F-theory models, this
symmetry originates from the gauge theory of seven-branes which intersect the
GUT\ model seven-brane. A potential refinement on the `diamond ring model'
which emphasizes the central role of $U(1)_{PQ}$ is given in section 7 of
\cite{HVGMSB}. This refinement is based on the observation that $U(1)_{PQ}$
naturally embeds in the group $E_{6}$ as the abelian factor of the maximal
subgroup $SO(10)\times U(1)_{PQ}$. Indeed, at the level of representation
theory, the $27$ of $E_{6}$ decomposes into irreducible
representations of $SO(10)\times U(1)_{PQ}$ as:%
\begin{align}
E_{6}  &  \supset SO(10)\times U(1)_{PQ}\\
27  &  \rightarrow1_{+4}+10_{-2}+16_{+1}\text{.}%
\end{align}
In this case, $X$ and the messenger fields respectively embed in the $1_{-4}$
and $10_{+2}$ of $\overline{27}$. In local F-theory models where the bulk
gauge group is given by $SU(5)$, a local enhancement in singularity type to
$E_{7}$ along a Riemann surface will contain matter fields in the $27$ of
$E_{6}$. This flexibility allows these local models to avoid much of the
baggage of four-dimensional GUT\ models with larger rank gauge groups.
Nevertheless, the construction presented in section 7 of \cite{HVGMSB}
contains some residual exotic matter fields so that the resulting matter
spectrum is only semi-realistic. Improving this type of construction is one
avenue of investigation which would be important to pursue in this approach.

\subsubsection{PQ Deformation of Gauge Mediation}

Leaving behind such aesthetic considerations, the matter content of the
effective theory generates an anomaly for the $U(1)_{PQ}$ gauge theory which
is canceled through a variant of the Green-Schwarz mechanism. As a
consequence, the $U(1)_{PQ}$ gauge boson develops a large mass through the
St\"{u}ckelberg mechanism. These results are well-known in the string theory
literature, but for further discussion on this point in the specific context
of these local F-theory models, see section 8.2 of \cite{HVGMSB}.

An important consequence of this fact is that heavy $U(1)_{PQ}$ gauge boson
exchange will generate a correction to the usual soft mass terms present in
gauge mediation scenarios. The precise value of this correction is fixed
by the mass of the gauge boson $M_{U(1)_{PQ}}$, and the fine structure constant
$\alpha_{PQ}$ of the $U(1)_{PQ}$ gauge theory so that at the messenger scale,
the soft masses squared for the Higgs fields and chiral matter receive an
additional correction of the form:%
\begin{equation}
m^{2}(M_{mess})=m_{GMSB}^{2}(M_{mess})+q\Delta_{PQ}^{2} \label{PQdef}%
\end{equation}
where $q=+2$ for the Higgs fields, $-1$ for all the other chiral superfields of the MSSM, and:%
\begin{equation}
\Delta_{PQ}^{2}=16\pi\alpha_{PQ}\left\vert \frac{F}{M_{U(1)_{PQ}}}\right\vert
^{2}\text{.}%
\end{equation}
Section 4.2 of \cite{HVGMSB} contains additional details on the derivation of
equation (\ref{PQdef}). When $M_{U(1)_{PQ}}$ is sufficiently low,
this effect constitutes an important, predictive deformation away from the
soft mass terms expected in the minimal gauge mediation scenario.

\subsubsection{$U(1)_{PQ}$ and the QCD Axion}

Remarkably, the phase of $X$ is also a viable candidate for the QCD axion. This
is because the $U(1)_{PQ}$ gauge symmetry is already Higgsed via the St\"uckelberg mechanism
at high energy scales, leaving behind a nearly exact global symmetry at lower energies.
The vev of the $X$ field breaks this nearly exact global symmetry, and the associated
Goldstone mode $a$ is the phase of $x$. As shown in section 6 of \cite{HVGMSB},
this phase directly couples to the QCD\ instanton density. The Lagrangian
density for $a$ includes the terms:%
\begin{equation}
L_{a}\supset\left\vert x\right\vert ^{2}\partial_{\mu}a\partial^{\mu}%
a+\frac{a}{32\pi^{2}}\varepsilon^{\mu\nu\rho\sigma}Tr_{SU(3)}F_{\mu\nu
}F_{\rho\sigma}\text{.}%
\end{equation}
This field mixes a small amount with other modes of the compactification. The
end result of this is that the axion decay constant is to leading order:%
\begin{equation}
f_{a}=\sqrt{2}\left\vert x\right\vert \sim10^{12}\text{ GeV}%
\end{equation}
which is within the standard allowed axion window with at most only mild fine tunings of
various couplings. Further background on axions and the potential role of the
phase of $x$ as the QCD axion may be found in section 6 and appendix A of
\cite{HVGMSB}.

\subsection{Explicit Supersymmetry Breaking Sector}

For the most part, the effects of supersymmetry breaking in the hidden sector
can be encapsulated entirely in terms of the vev of $X$. Nevertheless, for
completeness it is also important to develop models which realize the required
mass scales in a natural fashion. Section 8 of \cite{HVGMSB} provides an
explicit model of supersymmetry breaking based on a hybrid of a Fayet and
Polonyi model of supersymmetry breaking. Using the results of \cite{HMSSNV} on
instanton effects in the higher-dimensional anomalous $U(1)_{PQ}$ gauge
theory, it can be shown that a Polonyi-like linear term is generated:
\begin{equation}
W\supset M_{PQ}^{2}Q\cdot X
\end{equation}
where $Q$ is the instanton tunneling amplitude. For $M_{PQ}\sim M_{GUT}$ we
obtain $Q\sim5\times10^{-17}$. This contribution breaks supersymmetry and sets the
value of $F$ in equation (\ref{XVEV}) to the required value
$M_{PQ}^{2}Q\sim10^{17}$ GeV$^{2}$. The value of $x$ in equation (\ref{XVEV})
originates instead from the D-term potential of the anomalous $U(1)_{PQ}$
gauge theory due to a non-zero field dependent Fayet-Iliopoulos term $\xi_{PQ}$:%
\begin{equation}
V_{\text{Fayet}}=\left(  \left\vert x\right\vert ^{2}+...-\xi_{PQ}\right)
^{2}%
\end{equation}
where the \textquotedblleft$...$\textquotedblright\ refers to all other
contributions charged under $U(1)_{PQ}$. As shown in section 8 
of \cite{HVGMSB}, $\xi_{PQ}$ consists of a field dependent background value $\xi_{\ast}$ which
is always present in anomalous $U(1)$ theories, as well as another
contribution $\xi_{flux}$ from fluxes of the higher-dimensional theory through
the Riemann surface supporting $X$ so that:%
\begin{equation}
\xi_{PQ}=\xi_{flux}+\xi_{\ast}\text{.}%
\end{equation}
These two quantities are both large but will approximately cancel in a scan
over all available fluxes. The analysis of section 8.3 of \cite{HVGMSB}
establishes that the minimal non-zero value of $\xi_{PQ}$ is:%
\begin{equation}
\xi_{PQ}=\frac{M_{X}^{4}}{M_{pl}^{2}}%
\sim\left(  10^{12}\text{ GeV}\right)  ^{2}\text{.}%
\end{equation}
As a consequence, $\left\vert x\right\vert \sim10^{12}$ GeV, as required for
both gauge mediation and axion physics! We note that just as in the case of
neutrino physics reviewed in subsection \ref{NEUTRINOS}, the small hierarchy
$M_{GUT}/M_{pl}\sim10^{-3}$ is again crucial for realizing this effect.

\section{MSSM Parameter Space\label{ParamSpace}}

In this section we determine detailed features of the soft supersymmetry
breaking terms of the MSSM Lagrangian. The essential point is that the crude
considerations of previous sections on the scale of supersymmetry
breaking, the mediation mechanism, and the existence of a possible PQ
deformation are actually sufficient to completely fix the IR behavior of the
theory. While one might argue that this is simply a byproduct of the gauge
mediation scenario, there is a priori no reason to expect the
scale of supersymmetry breaking to be compatible with this scenario, or for
various details of the messenger sector to be constrained at even a crude
level, as we have done here.

To ensure compatibility with electroweak symmetry
breaking, and to determine the sparticle spectrum, in
\cite{HVGMSB} we utilized the program \texttt{SOFTSUSY}
\cite{SOFTSUSYAllanach}. After specifying some details of the messenger
sector, this program adjusts the IR boundary conditions of
the theory to remain in accord with electroweak symmetry breaking. These
further considerations fully determine the remaining UV boundary conditions,
which in turn constrain the remaining parameters of the MSSM.

Some sample scans over various regions of the constrained UV boundary
conditions are provided in Section 9 of \cite{HVGMSB}. As representative
examples, we mainly considered models with a single vector-like pair of
messenger fields in the $5\oplus\overline{5}$ of $SU(5)$. Because small changes in the
messenger scale $M_{mess}$ only altered the resulting sparticle masses by a
small amount, we took as fixed the value $M_{mess}=10^{12}$ GeV. The scans
presented in \cite{HVGMSB} primarily focused on the two parameter subspace
defined by $\Lambda=F/x \sim 10^{5} - 10^{6}$ GeV and the PQ deformation
$\Delta_{PQ} \sim 0 - 10^{3}$ GeV. The final input necessary for
determining the UV boundary conditions is given in terms
of $\tan\beta \equiv \langle H_u \rangle / \langle H_d \rangle$ at the
messenger scale. In order to remain in accord with
electroweak symmetry breaking, \texttt{SOFTSUSY} adjusts the UV values of
$\mu$ and $B\mu$. Scanning over a range of values for
$\tan\beta$, it is then
possible to recover the boundary condition $B\mu=0$ at the messenger scale. We find
that at the weak scale, $\tan\beta \sim 30 \pm 7$, where the particular
numerical value depends on the specific UV boundary conditions imposed. While
perhaps obvious, note that a large $\tan \beta$ scenario is quite compatible with
the expectation that in GUT models, the Yukawa couplings for up and down type quarks
are expected to be comparable, order one numbers. In such a situation, achieving a
large hierarchy between the mass of the bottom and top quark, for example, requires that the vev
$\langle H_{u} \rangle$ be at least an order of magnitude larger than $\langle H_{d} \rangle$.

\begin{figure}
[ptb]
\begin{center}
\includegraphics[
height=4.12in,
width=6.4091in
]%
{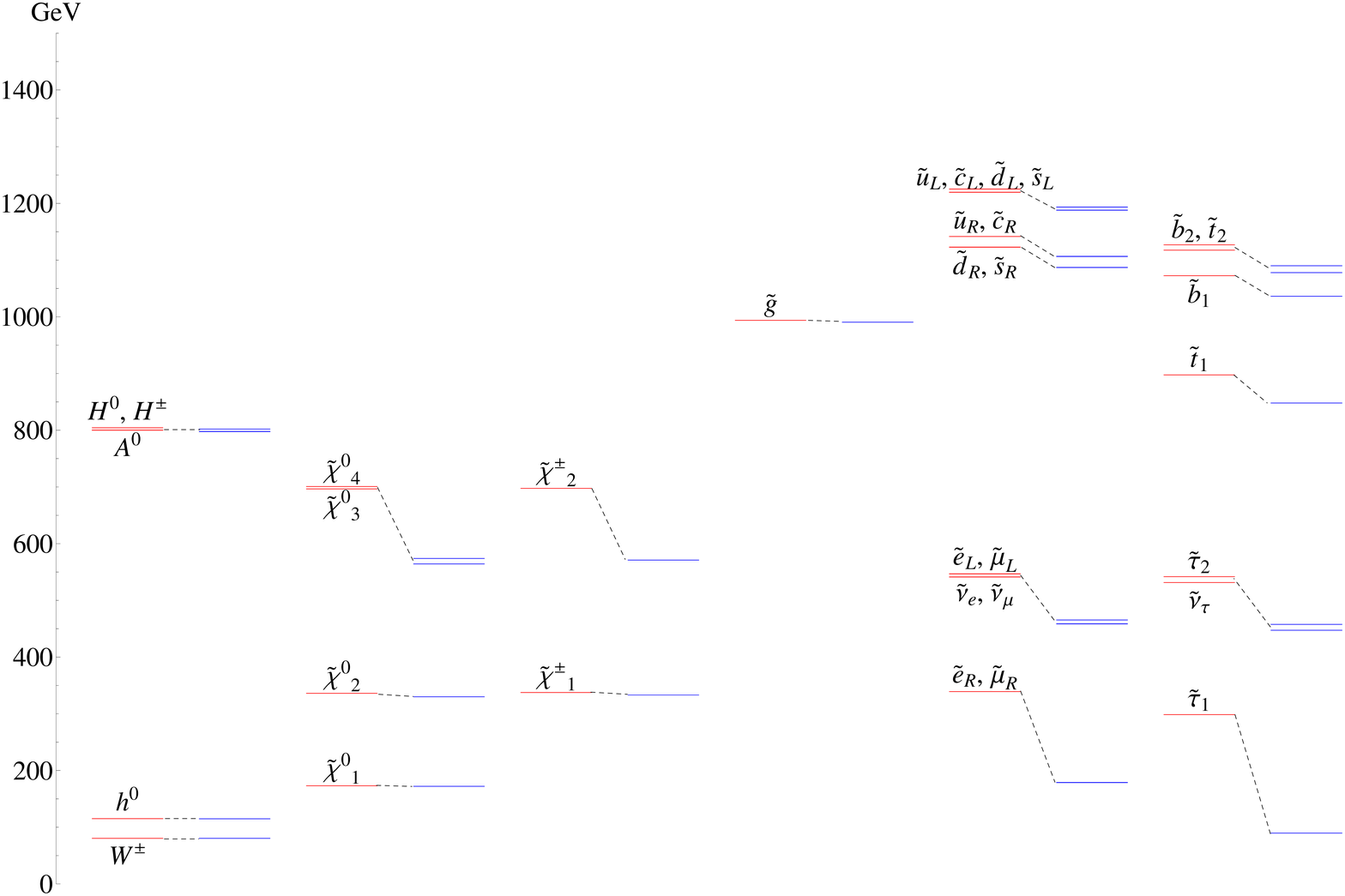}%
\caption{Plot originally presented in \cite{HVGMSB} of the sparticle spectrum
(sparticles denoted by a \ $\widetilde{}$ ) in a gauge mediation scenario with $\Lambda=1.3\times10^{5}$ for a single
vector-like pair of messenger fields in the $5\oplus\overline{5}$ of $SU(5)$
at vanishing PQ\ deformation (red), and for a maximal PQ\ deformation of
$\Delta_{PQ}=290$ GeV (blue). Beyond this point, a tachyon is present in the
squark/slepton sector. This deformation causes the lightest stau
$(\widetilde{\tau}_{1})$ to become the NLSP. Note also that at large PQ
deformations, the selectron and smuon $(\widetilde{e}_{R},\widetilde{\mu}%
_{R})$ are comparable in mass to the bino-like lightest neutralino
$(\widetilde{\chi}_{\text{ \ }1}^{0})$.}%
\label{together}%
\end{center}
\end{figure}

As in nearly all gauge mediation models, the lightest supersymmetric particle (LSP)
is the gravitino, and in our case its mass is $\sim 10-100$ MeV. Scanning
over the UV boundary conditions reveals that for the most part, the bino is
indeed the next lightest supersymmetric particle (NLSP). Large values of the PQ
deformation can sometimes alter this story for sufficiently low values of $\Lambda$
by allowing the stau to decrease in mass to the point where it becomes first a
co-NLSP with the bino, and then the NLSP. There are
limits to how large the PQ deformation can become, because it will eventually
cause the scalar effective potential to develop a tachyonic mode, other than
the one present in the Higgs sector. Figure \ref{together} shows a plot of the sparticle masses in a
scenario with a low value of $\Lambda$ which is consistent with the current
bound on the mass of the Higgs for vanishing PQ deformation, and for a maximal
stable PQ deformation. In this case, the lightest stau can become the NLSP.
For larger values of $\Lambda$, the PQ deformation induces an instability in
the squark/slepton effective potential before the stau can become the NLSP.
Many further details on the sparticle spectrum, as well as a discussion on the
amount of fine-tuning in the Higgs sector can be found in section 9 of
\cite{HVGMSB}.

\subsection{Discovery Potential at the LHC}

The rigid framework we have described in previous sections makes definite
predictions for what should be seen at the LHC. That being said, given a
particular experimental signal, it is a notoriously difficult problem to
reconstruct from LHC data alone a given model of beyond
the Standard Model physics. In this regard, it is perhaps more important to determine
signals which could \textit{falsify} this class of models.

In many gauge-mediated supersymmetry breaking scenarios, the NLSP\ is either a
bino-like lightest neutralino, $(\widetilde{\chi}_{\text{ }1}^{0})$, or the
lightest stau $(\widetilde{\tau}_{1})$. Here, the \ $\widetilde{}$ \ indicates
that these are sparticles. These sparticles will decay to the
gravitino LSP\ through the respective processes $\widetilde{\chi}_{\text{ }%
1}^{0}\rightarrow\widetilde{G}_{3/2}\gamma$ and $\widetilde{\tau}%
_{1}\rightarrow\widetilde{G}_{3/2}\tau_{1}$. Although this is a gauge
mediation scenario, the high scale of supersymmetry breaking implies that the
NLSP\ decays outside of the detector, effectively behaving as an\ LSP. As
reviewed for example, in \cite{MartinPrimer,GiudiceSUSYReview}, the average decay length
for an NLSP\ with mass $m$ produced with energy $E$ is:%
\begin{equation}
L=\frac{1}{\kappa_{\gamma}}\left(  \frac{m}{100\text{ GeV}}\right)
^{-5}\left(  \frac{\sqrt{F}}{100\text{ TeV}}\right)  ^{4}\sqrt{\frac{E^{2}%
}{m^{2}}-1}\times10^{-2}\text{ cm}\text{,}%
\end{equation}
where in the above, $\kappa_{\gamma}$ is a constant which depends on details of the NLSP. In
a crude approximation, we can essentially set all of the above factors to
unity except for the term involving $\sqrt{F}$. Because $\sqrt{F}\sim10^{8.5}$
GeV, the resulting decay length is:%
\begin{equation}
L\sim10^{12}\text{ cm,}%
\end{equation}
which is well outside the particle detector.

The consequences of this for the LHC depend on whether the $\widetilde{\chi
}_{\text{ }1}^{0}$ or the $\widetilde{\tau}_{1}$ is the NLSP. When the
$\widetilde{\chi}_{\text{ }1}^{0}$ is the NLSP, it will simply leave the
detector as a neutral particle. This is in sharp contrast to many models of
gauge mediated supersymmetry breaking where the scale of $\sqrt{F}$ is
significantly lower. Indeed, one striking prediction of many gauge
mediation scenarios are events with two hard photons from the decay of $\widetilde{\chi
}_{\text{ }1}^{0}$'s, which is clearly \textit{not} the case in the present class of models.

When the $\widetilde{\tau}_{1}$ is the NLSP, we can expect events with two
charged tracks through the detector calorimeter. Due to the large difference
in mass between these particles and the muon, it is then possible to
distinguish this signature from a generic muon event.\footnote{For an extensive study
of a stau NLSP in the related, although ultimately different, sweet spot
model of supersymmetry breaking, see \cite{KitanoIbeSweetSpot}. We caution,
however, that the results of this reference are not directly applicable to the
case at hand, because the sparticle spectrum is somewhat different
there. For example, the heaviest neutralino $(\widetilde{\chi
}_{\text{ }4}^{0})$ is primarily a wino in \cite{KitanoIbeSweetSpot}, although
in the present class of models it is a higgsino. Some other differences
include the mass of the gravitino, which is typically heavier at around $1$
GeV, than in the models we consider.}

\begin{figure}
[ptb]
\begin{center}
\includegraphics[
height=1.7685in,
width=2.2182in
]%
{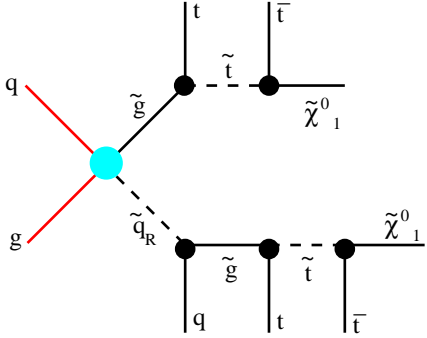}%
\caption{Depiction of a sample decay chain in a scenario where the bino-like
lightest neutralino $(\widetilde{\chi}_{\text{ }1}^{0})$ is the NLSP. Starting
from the initial collision of partons, the end result is two $\widetilde{\chi
}_{\text{ }1}^{0}$'s and some number of top quarks and anti-quarks. These tops will then
decay further into jets which will sometimes also include leptons in
the end result.}%
\label{decaychain}%
\end{center}
\end{figure}

Using the program \texttt{PYTHIA} \cite{PYTHIA}, we have determined the cross
sections for such high scale gauge mediation models.\footnote{We thank T.
Hartman for very helpful explanations on how to use \texttt{PYTHIA}.} Regardless of the particular
details of the NLSP, it appears likely that the LHC will be able to produce a
sufficient number of events to allow some crude features of these models to be
either verified or falsified. For example, in the scenario considered earlier
with a single vector-like pair of messenger fields, $\Lambda
=1.3\times10^{5}$ GeV, and $\Delta_{PQ}=0$, the mass of the gluino is
$\sim1000$ GeV, and the mass of the lightest stop is $\sim900$ GeV. In this
case, the MSSM\ process with the largest total cross section is the quark
gluon parton collision $qg\rightarrow\widetilde{q}_{R}\widetilde{g}$. We
find that $\sigma\left(qg\rightarrow\widetilde{q}_{R}\widetilde{g}\right)  \sim3\times10^{2}$ fb.
See figure \ref{decaychain}\ for a depiction of one decay chain of
$qg\rightarrow\cdot\cdot\cdot\rightarrow \slashed{E}_{T}+jets$ which has a relatively
large branching ratio at each vertex. Here, $\slashed{E}_{T}$ denotes missing transverse energy. A similar analysis can also be
performed for models with a maximal PQ deformation turned on, although in this case
the appearance of tracks in the calorimeter is likely to be a more reliable
tool for discrimination. It is beyond the scope of this paper to present a
more complete analysis of potential collider signatures. Indeed, while certain
processes may have large cross sections, it is likely that signatures with
less QCD background could be of greater utility. It would be potentially quite interesting to go beyond the quick
sketch presented here to investigate this set of issues in detail.%

\section{Conclusions and Future Directions\label{CONC}}

In this paper, we have given a short overview to our recent work on
constructing GUT models in F-theory. One of the perhaps surprising outcomes of
this analysis is that simple, qualitative considerations can have far-reaching
consequences on both the UV, and IR behavior of the theory. This, rather than
any particular mechanism endows these models with surprising predictive power.
While we have reviewed the main threads of analysis which lead from the
GUT\ scale all the way down to the weak scale, there are many additional
ingredients which we have only briefly touched on, and which are covered in far
greater detail in the three papers \cite{BHVI,BHVII,HVGMSB}. We have also
presented a short description of potential ways that the LHC could discover
evidence either in favor of, or against, these types of models.

There are potentially other ways that the ingredients described above could fit
together to form a viable phenomenological model. One example would be to find
refinements to our solution to the $\mu/B\mu$ problem in gauge mediation
models based on embedding the $U(1)_{PQ}$ symmetry into an $E_{6}$ GUT. In
this class of models, the exact spectrum of the MSSM remains to be
constructed, but further analysis of the associated geometries provides a
potentially promising avenue of investigation.

Another important ingredient is the way that the small hierarchy between the
GUT\ scale and Planck scale $M_{GUT}/M_{pl}\sim10^{-3}$ has appeared
repeatedly throughout these F-theory models, and especially in
\cite{BHVII,HVGMSB}. In the most predictive models we have found, this mild
hierarchy appears in the form of some power law dependence, which tethers the physics of
the axion, neutrinos, and messengers to the scale $10^{12}$ GeV. There
is also the possibility that a stronger hierarchy could be generated due to exponential suppression of wave
functions near the GUT model seven-branes. By its nature, these effects lead
to less predictive models, but they also provide additional flexibility, and
this topic would be interesting to study in further detail.

Finally, the primary focus of the work presented in this paper has been on
potential realizations of particle physics models. Cosmological constraints
constitute another important avenue of investigation \cite{CosmoStudy}, which
are likely to shed further light on details of both the UV and IR regimes of
the theory.

\section*{Acknowledgements}

We especially thank C. Beasley for a very productive collaboration in
\cite{BHVI,BHVII}. In addition, we also thank B.C. Allanach, B. Andreas, K.S.
Babu, V. Bouchard, F. Denef, A.L. Fitzpatrick, G. Giudice, T. Hartman, J.
Marsano, D.R. Morrison, S. Raby, N. Saulina, S. Sch\"{a}fer-Nameki, P.
Svr\v{c}ek, A. Tomasiello, A.M. Uranga, H. Verlinde, M. Wijnholt, E. Witten,
and S.-T. Yau for helpful discussions. We would also like to thank the Fifth
and Sixth Simons Workshop in Mathematics and Physics for hospitality while
some of this work was performed. JJH would also like to thank the 2008
Amsterdam Summer Workshop on String Theory for hospitality while some of this
work was performed. The work of the authors is supported in part by NSF grant
PHY-0244821. The research of JJH was also supported in part by an NSF Graduate Fellowship.

\bibliographystyle{ssg}
\bibliography{fgutsgmsb}

\end{document}